\documentclass[a4,12pt]{article}
\usepackage{graphicx}
\usepackage{graphics}
\usepackage{epsf,epsfig}
\usepackage{amsmath}
cm \textheight 25 true cm \topmargin=-2mm

\begin{document}
\thispagestyle{empty}\setcounter{page}{1}
 \vskip40pt
\centerline{\bf Non Markovian Noise mediated through Anamolous Diffusion within Ion Channels} \centerline{\bf }

\vskip10pt

\centerline{\footnotesize Sisir Roy $^{1,2}$, Indranil Mitra$^3$, Rodolfo Llinas$^4$}
\vskip5pt 
\centerline{\footnotesize$^{1}$Physics and
Applied Mathematics Unit, Indian Statistical Institute,}
\centerline{\footnotesize 203 Barrackpore Trunk Road, Kolkata
700108, India}

\centerline{\footnotesize$^{2}$College of Science,
George Mason University} \centerline{\footnotesize 4400 University
Drive Fairfax, Virginia 22030,USA}
\vskip5pt
\centerline{\footnotesize$^{3}$ Brain \&\ Behaviour Program,
Department of Physics, Georgia State University}
\centerline{\footnotesize 29 Peachtree Center Avenue, Atlanta,GA
30303 USA }
\vskip5pt

\centerline{\footnotesize$^{4}$ New York University School of Medicine, 530 First Avenue, New York}
\centerline{\footnotesize NY, 10016 USA}

\vskip20pt
\begin{abstract}
It is quite clear from a wide range of experiments that gating
phenomena of ion channels is inherently stochastic. It has been
discussed using BD simulations in a recent paper that memory
effects in ion transport is negligible, unless the barrier height
is high. In this brief report we like to state using Differential Stochastic Methods (DSM's) that
the Markovian property of exponential dwell times do indeed give
rise to a high barrier, which in turn indicates that memory
effects need not be ignored. We have thus constructed a Generalized Langevin Equation which
contains a combination of Non Markovian at different time scales
\&\ Markovian processes and develop an algorithm to describe the
scheme of events. We see that the oscillatory function behaviour
with exponential decay is obtained in the Markovian limit and two
distinct time scales corresponding to the processes of diffusion
\&\ drift may be obtained from preliminary simulation results. We
propose that the results need much more inspection and it will be
worthwhile to reproduce using MD simulations. The most important idea which we like to propose in this paper is that the rise of time scales and memory effects may be inherently related to the differential behaviour of shear viscosity in the cytoplasm \&\ extracellular matrix.

\end{abstract}

\newpage

\section{Introduction}

Ion channels are like gates which facilitate exchange
of electrolyte between the exterior and interior of a cell.
Pores are formed by specific proteins embedded
into the phospholipid membrane.
Depending on the conformation of the
protein, the pore can be open or closed. When open,
the protein is
very specific to the kind of ions that it allows to pass through
the channel. Ion channels are usually
very narrow and pass through a region of low dielectric constant, the
lines of electric field tend to be confined to the high dielectric
interior of the pore.
One
can easily see this by considering the requirement of the continuity of the
normal component of the electric
displacement field between the interior of a channel and membrane, $\epsilon_w {\bf E}_w^n=\epsilon_p {\bf E}_m^n$.  Since
the lipid membrane has a dielectric constant $\epsilon_p \approx 2$ while
the dielectric constant of water is $\epsilon \approx 80$ we immediately
see that the normal component of the electric field at the membrane/pore
boundary must be very close to zero.
There is only a very slight
penetration of
the electric field into the interior of the phospholipid membrane.
The situation here is, therefore, very
similar to the expulsion of the magnetic field
by a superconductor.
For example, for a channel of length $L=25$ \AA
$\,$ and radius $a=3$ \AA, the barrier is about $6k_BT$,  which although
still quite large, should allow ionic conductivity.
Recently the study  of ion channels
has expanded to other parts of applied physics.  Water filled
nanopores are introduced into silicon oxide films, polymer membranes,
etc.\cite{int01,int02}.

Many methods ranging from Molecular (MD) \&\ Brownian dynamics (BD) simulations with implicit water treated as a uniform dielectric continum, to the mean field Poisson-Nerst Planck Equation (PNP), which treats both the ions \&\ water implicitly
have been used to simulate the ion channel phenomena charecterstics.
While clearly the most accurate, MD simulations are computationally very
expensive\cite{int03}.
Brownian dynamics is significantly
faster than MD, but because of the dielectric discontinuities across the
various interfaces a new solution of the Poisson equation
is required for each new configuration of ions inside the pore.
The simplest approach to study the ionic conduction
is based on the PNP theory\cite{int04,int05}. This combines the continuity
equation with the Poisson equation and Ohm's and  Fick's laws.
PNP is intrinsically mean-field and is, therefore, bound to fail when
ionic correlations become important.

For narrow channels, the cylindrical geometry, combined with the
field confinement, results in a pseudo one dimensional potential of
very long range\cite{int06}.
Under these conditions the correlational effects dominate, and
the mean-field approximation fails\cite{int07}.
Indeed recent comparison between the
BD and the PNP showed that PNP breaks down when the pore radius is smaller
than about two Debye lengths\cite{int08,int09}.
At the moment, therefore, it appears that
a semi-continuum (implicit solvent) Brownian dynamics simulation is
the best compromise between the cost and
accuracy\cite{int10,int11,int12} for narrow pores.
If the interaction potential
between the ions inside the channel would be known, the simulation
could proceed orders of magnitude faster.

It is worthwhile to mention here that though BD simulations of ion channels have yielded suitable results,
but most importantly it has been confined with the use of Langevin equation with Markovian random forces.
The Markovian approximation is justified when the Brownian particles are much heavier than the solvent molecules,
 a condition which is obviously not satisfied for ions in water. Correlations are important and can be taken into account by
 Generalized Langevin Equation (GLE)\cite{int13}. In a recent paper these issues have been discussed in the context of memory effect and it is
 found that memory effect are negligible in bulk simulations of electrolytes except that when ions have to cross energy barriers, inclusion of the
 memory effects via the GLE becomes important. We argue here that interesting physics of ion channels are concentrated in the pore region where the
 stochasticity of gating phenomena is crucial for the anaysis. We argue here by the Einstein result
 $\sigma=nq^{2}D/kT$, that the ionic conductivity exhibits a dependence
 on $\Delta t$ which may be employed in the BD simulations at
 lower field strength, which disappears for higher field. But as
 we have mentioned before that the phenomena being stochastic the
 potential energy barrier has also a stochastic contribution with
 the barrier height being a random variable as $ V(z,i)=
 V_{i}exp(-z^{2}/2d^{2})$ where $V_{i}$ is the random variable
 associated with the barrier height and $d$ is the width. So in contrary
 to the recent survey \cite{int14} memory effects will be important in a
 variety of cases depending on the random effects of the potential
barrier.

The most important part of the understanding of the dynamic aspects of the memory effects can be understood by an insight driven by some recent
development in Atomic Force Microscopy which suggests the capillary condensation of water, which increases sort term memory in the form of shearing
elastic frictional forces. It suggests that at the nanoscale there is a crucial change in the elastic response of the forces between water molecules
which leads to a variable viscosity in the cytoplasm in comparison to the $K^{+}$ channels which is the key process in giving rise to memory effects.
What however needs a serious investigation is how the shear response behaviour may be exactly coupled with the Stochastic dynamics.
In this context, we would like to mention that diffusion governed by Brownian motion is an efficient transport mechanism on short time and length scales. Even a highly organized system like a living cell relies in many cases on the random Brownian motion of its constituents to fulfill complex functions.
A Brownian particle will rapidly explore a heterogeneous environment that in turn strongly alters its trajectory. Thus, detailed information about the environment can be gained by analyzing the particle's trajectory. It is well known that the non-negligible fluid's inertia leads to hydrodynamic memory effects \cite{int15} resulting in a characteristic long-time tail of $t^{-3/2}$ in the velocity correlations of the particle's motion.
The viscosity of the PE solution is recently calculated \cite{int16} as a function of the concentration and characterize hydrodynamic memory effects
present is such solution. We find that memory is lost above the critical concentration C*, which corresponds to the point where neighboring
 polymer coils start to overlap and form a transient mesh where fluid dynamics is dominated by viscous terms rather than inertial ones.
 Above this point the mesh surrounding the Brownian particle increases its effective mass and therefore the characteristic power law of
 the velocity correlations vanishes.
In this paper we construct a GLE including Non Markovian
processes using an white noise a linear combination of colored
noises, in the spirit of Ornstein-Uhlenbeck process. Our model
includes an evolution of memory and the mixed process generates
oscillations within the pore. For purely Markovian or purely Non
Markovian processes the time dependence is monotonic. So it is
clear that the oscillations are the result of the competitive
interactions between the Markovian and Non Markovian
subprocesses. The non Markovian nature of the noise effects both
the time dependent and stationery properties of the driven
stochastic processes. The time dependent processes are
charecterized by distinct time scales. In the next section we
give a justification of time scales for a diffusive process and
thereafter describe our model and a specific algorithm for the
mixture of the non-Markovian and Markovian processes.

\section{Structure of Potassium Channels}

A lot of experimental effort has been put into re-solving membrane-bound processes, from the structural as well as dynamical point of view. As far as the molecular structuredetermination of membrane proteins is concerned,obtaining a cristallized functional form of the membrane protein is still a generally non-solved problem,up to now, only few structures have been resolved upto the atomic details. Consequently, modelling and simulations of membrane processes has been long facing the problem of lack in structural data, not to mention the problem of computational limitations for such a complex systems. The advantage of the simulation over experimental techniques has been widely seen in the possibility to explore the dynamical aspects of the structure which cannot be adressed experimentally. At present, the amenable combination of atomistic resolution structures with highly sophisticated computational methods is considered to be a leading way to better understanding of fundamental properties of basic cell membrane physiology. The doubts about  physiological relevance of results obtained on bacterial membrane proteins to those in mammalian cells have been lifted at least in the case of KcsA,for which there is evidence supporting evolutionary conservation of the architecture as well as of transporting properties. The membrane resting potential, which is predominantely determined by K+concentration difference, can easily be changed by allowing the Na+ions flow inside; if the membrane is depolarized, the resting potential is effectively restored by letting the K+flow out whereas subtle changes in Ca2+ intracellular concentration can lead to a number of processes affecting important cell functions, etc.Ion channels are transmembrane protein struc-tures \cite{sec101} which ensure the continuous transportpathway for diffusive flow of ions accros the mem-brane. As the direction of the ion flow through thechannel is down the electrochemical gradient for theion, the process is often reffered to as a passive trans-port. Since the net flow of electric charge gives rise to a rapid change in transmembrane potential, ion channels play a key role in generating and propagating action potentials in nervous system. Moreover,it is well established that ion channels play important role in pathophysiology of various diseases and thereby present the primary targets for pharmacological drug design.In order to assure the precisely controlled transmembrane ionic flow, ion channels have to be highly selective towards specific ion type and have to be well defined gating control mechanism. The protein conformation of the open state, in which the selected ions are allowed to pass, has to be accomplished upon specific stimulus, the known mechanisms include change in transmem-brane potential, bindingof another molecule and mechanical stress.Although the experimentally determined rate of ion transport through different channels varies, the simple calculation in given circumstances reveals that this rapid transport is almost at the diffusion rate\cite{sec102}. This fact is indicates that an open channel structure is ensuring energetically almost barrierless pathway for selected ion flow. Additionally, the variability among channels is also seen in the geometry of ion transport pathway, but there seem to be two characteristic regions appearing as a general feature of all biological channels, a wider cavity that accomodates hydrated ions and a shortand narrow selectivity filter. Moreover, those two regions seem to be evolutionary conserved, at least in specific channel type.Ion channels are functionally and structurally simplest among the membrane transport proteins, sincethe transport itself doesn't call for a conformational changes of protein structure. Conformational changes occur in the process of gating.There are three major experimentally determined descriptors of channel function : 1) high selectivity towards one ion species 2) permeation process that ensures rapid transport at given rate, defining the conductance of the channel (I-V curves of the ma-jority of biological channels are linear within the physiological range of membrane potentials), and 3)gating mechanism.

Potassium channels are membrane-spanning proteins that provide an energetically favorable pathway for the selective conduction of K+ ions across the membrane. One of the most striking properties of potassium channels is their remarkable ability to conduct K+ ions near the diffusion limit and yet, select for K+ over Na+ by more than 1000 to 1. Because small ions such K+ are strongly bound to water molecules in bulk solution, the channel provides coordinating groups that help compensate the loss of hydration. Selectivity arises when this energetic compensation is more favorable for one type of ion than for another, relative to the hydration free energy. The most relevant ions in biological systems are Na+, K+, Cl-, and Ca2+, but Na+ and K+ are the most abundant, with a high intracellular concentration for K+ and a high extracellular concentration for Na+. The molecular mechanism underlying the rapid discrimination between K+ and Na+ is, therefore, fascinating because these two monovalent cations are very similar, differing only slightly in their atomic radius. Assuming that ions surrounded by complete shells of water molecules would maintain monotonic and circular profiles, he argued that such a cylindrical channel might select a partially hydrated ion of an optimal size while discriminating over smaller or larger ions: If K+ approaches a pore that is precisely the same size as this ion with its first solvation shell, it may, as indicated previously, exchange hydration, for water shells to infinity, for a similar attraction with the structure lining the pore. If the pore is somewhat smaller that K+ penetration cannot occur for steric reasons, while if pore is somewhat too large, penetration likewise cannot occur because the attraction of the ion for water shells and greater is not compensated by solvation of similar magnitude in the pore.

The determination of the three-dimensional structure of K+ channels at atomic resolution using X-ray crystallography provides  the tools to deepen our understanding of these systems. In the narrow selectivity filter, K+ must be almost completely dehydrated. These observations led to a commonly accepted explanation of ion selectivity, which assumes that structural factors play the dominant role. Potassium channels, structurally, are membrane proteins that have a signature selectivity filter in a "pore region" that is flanked by transmembrane helices on either side \cite{sec103}. Together, this structural motif constitutes the permeation pathway of the channel and subfamily designations commonly indicate the attachments to this basic design. The pore-forming monomers are usually referred to as alpha-subunits and most often form homotetramers in the membrane in order to function. It is reasonable to suppose that the diverse functional properties of potassium channels necessarily stem from the nature of their particular sequence \cite{sec103}. Hydropathy plots, which can be used to infer transmembrane (TM) topology, are a measure of potassium channel diversity. Potassium channels with a 8 TM-2 pore topology have also been discovered in yeast \cite{sec104,sec105}. A description of the expansion of K channel subfamilies has been provided in higher animals based on topology and sequence similarity of many sequences. The physiological diversity of potassium channels places a significant bottleneck to formulating a welldelineated classification scheme for this protein family.

The structure of the KcsA channel, is strikingly consistent with the classical views of a very selective, fast-conducting, multi-ion pore. The pore comprises a wide, nonpolar aqueous cavity on the intracellular side, leading up, on the extracellular side, to a narrow pore that is 12`` long and lined exclusively by main chain carbonyl oxygens. Formed by the residues corresponding to the signature sequence TTVGYG, common to all K+ channels , this region of the pore acts as a selectivity filter by allowing only the passage of nearly dehydrated K+ ions across the cell membrane. The X-ray structure unambiguously demonstrated that the K+ ion entering the selectivity filter have to lose nearly all their hydration shell and must be directly coordinated by backbone carbonyl oxygens.  The filter is constrained in an optimal geometry so that a dehydrated K+ ion fits with proper coordination but the Na+ ion is too small, in close correspondence with the snug-fit mechanism of Bezanilla and Armstrong \cite{sec106}. This simple and appealing structural mechanism was then widely adopted to explain the selectivity of the K+ channel, as explicitly stated by several works.

\begin{itemize}
\item A rigid K+ pore, however, cannot close down around a Na+ ion, which does not bind snugly in the pore and thus has a much higher energy than in water.
\item Each K+ ion in the selectivity filter is surrounded by two groups of four oxygen atoms, just as in water: these oxygen atoms are held in place by the protein, and are in fact the backbone carbonyl oxygens of the selectivity filter loops from the four subunits. Furthermore, they solve the problem of stabilizing potassium in preference to sodium by precisely matching the configuration of oxygen atoms around a solvated potassium ions.
\item The filter, for structural reasons, cannot constrict sufficiently to bring more than two of the carbonyls within good bonding distance of the Na+. As a result, the energy of the Na+ in the pore is very high compared with its energy in water\cite{sec107}.
\item Potassium fits optimally at these sites. While they expand to accommodate rubidium and cesium, they don't contract enough to cradle sodium.
\item The channel pays the cost of dehydrating K+ by providing compensating interactions with the carbonyl oxygen atoms lining the selectivity filter. However, these oxygen atoms are positioned such that they do not interact very favorable with Na+ because it is too small. Because of its relative rigidity the channel would not afford favorable interaction with ions of with different than potassium radius.
\end{itemize}

The main idea is that the narrow pore is perfectly suited (at the subangstrom level) to provide a cavity of the appropriate size to fit K+, but unable to adapt to the slightly smaller Na+. This implies a significant structural inability to deform and adapt: the energetic cost upon collapsing to cradle a Na+ (a structural distortion of about 0.38 angstroms) must give rise to a significant energy penalty (much larger than $k_{B} T$ assuming the existence of molecular forces opposing a sub-angstrom distortion is tantamount to postulating structural rigidity. Furthermore, the geometry of such a rigid pore must be very precisely suited for K+ because it would be unable to adapt small perturbations without paying a significant energy price (much larger than $k_{B}T$). Therefore, precisions in structural rigidity and geometric precision are two underlying microscopic consequences to these.
However there are fundamental problems with the common view. Proteins, like most biological macromolecular assemblies, are soft materials displaying significant structural flexibility\cite{sec108}. Despite some uncertainties, the B-factors of the KcsA channel indicate that the RMS fluctuations of the atoms lining the selectivity filter are on the order of 0.75 to 1.0\AA{A} , in general agreement with numerous independent MD simulations of KcsA. The magnitude of atomic thermal fluctuations is fundamentally related to the intrinsic flexibility of a protein, i.e., how it responds structurally to external perturbations. These considerations suggest that, at room temperature, the flexible/fluctuating channel should distort easily to cradle Na+ with little energetic cost, as is seen in MD simulations with Na+ in KcsA. The flexibility of the pore is further highlighted by the experimental observation that K+ is needed for the overall stability of the channel structure\cite{sec110,sec111}. Therefore, even ion channel proteins appear to be inherently too flexible to satisfy the requirement of the traditional snug-fit mechanism. Furthermore, structural flexibility is absolutely essential for ion conduction since in some places the diameter of the pore in the X-ray structure of KcsA  is too narrow to allow the passage of a water molecule or a K+ ion.

In the electric circuit equivalent the inserted membrane proteins thereby play the role of field-effect transistors, with a voltage imposed across the cell membrane 'gating' the transfer of ion bound charges through the membrane. Two different aspects characterize channel function: ion-selective permeation and 'gating', i.e. control of access of ions to the permeation pathway. We will base the subsequent concept on potassium channels, employing the crystal structure of the KcsA and KvAP channels at a resolution ranging from 1.9x 10-10 m to 3.2 x 10-10 m, as revealed by the work of  MacKinnon's group. The channel structure is basically conserved among all potassium channels with some differences relating to gating characteristics rather than ionic selectivity. In the open gate configuration the protein selects the permeation of K+ ions against other ions in the selectivity filter'and can still allow ion permeation rates near the diffusion limit. In the view of HH-type models of membrane potentials K+ permeation stabilizes the membrane potential, resetting it from firing threshold values to resting conditions. The atomic level reconstruction of parts of the channel and accompanying  molecular dynamics simulations (MD) at the 10 -12 sec resolution have changed the picture behind ion permeation: the channel protein can transiently stabilize three K+ states, two within the permeation pathway and one within the 'water-cavity' located towards the intracellular side of the permeation path. 
The essential feature behind the channel structure relating to the present work is provided by the closed gate, low-K+ state of the permeation pore of the protein, to which the original crystallographic image. Whereas in the open gate state, with the cavity exposed to the high  intracellular K+ concentration, the interior of the protein represents an almost barrierless pathway for selected ion flow,  the closed gate state represents a stable ion-protein conformation.

More recently, the development of computational approaches based on sophisticated all atom molecular dynamics simulations\cite{sec113} has started to offer a virtual route for testing various ideas about the molecular mechanism of ion selectivity. One of our goals with this work, in addition to review the recent results from modern computations, will be to provide a model based on the relation of Non markovian processes linked with a proposal of different viscosity of water inside \&\ outside the channels to get away with the pore problem.

\section{Gating Kinetics \&\ the Role of Diffusion}

The equation describing diffusive motion in a 1D potential of
mean force (pmf) is given by the Smoluchowski equation as
\begin{equation}\label{smol1}
  \dot{p}(z,t)= -\frac{\delta}{\delta z}\Large[-\frac{V'(z,i)}{R(z)}p -\frac{kT}{R(z)}\frac{\delta}{\delta
  z}p ]\Large
\end{equation}
where $p$ is the probability density of the gating particle,
$V(z,i)$ is the random pmf and $R(z)$ is the friction
coefficient. The equation mainly states the local probability
conservation corresponding to the processes of drift and
diffusion. Drift motion occurs in the presence of pmf gradient
producing a directed displacement of the probability distribution
to a a region of local energy minimum. Diffusive motion allows
activated transitions over energy barriers to take place. In
equilibrium drift and diffusion balance each other through
Boltzman distribution. Well defined states happen only for local
minima in the energy landscape and to satisfy the Markovian
property of exponential dwell times, activation barriers must be
sufficiently high. So in accordance to the previous analysis we
see that in these cases the role of GLE becomes inevitable and we
need to understand the dynamics of ion channels through GLE by
extending the process to be an admixture of both Markovian and
Non Markovian processes. Experimentally drift phenomena may be
measured by large bandwidth recordings of the initial transient
current after a rapid voltage jump. Evidence for the existence of
an early fast component of gating comes with the recent
observation  obtained in gating currents of giant patch
recordings of K channels. Apart from these there is another sow
physiological time scale, which is characterized by the dwell
time within the states, and can be measured by the detection of
fluctuations in the gating current which has`the`features of shot
noise.

Equation \ref{smol1} is of the form of a Fokker-Planck(FP)
Equation as

\begin{equation}\label{smol2}
\frac{\partial p}{\partial t} = F p
\end{equation}

where $ F= -\frac{\delta}{\delta z} A(z)
+\frac{1}{2}\frac{\delta^{2}}{\delta z^{2}}B(z)$ and $A, B$ are
the operators constructed from \ref{smol1}. The energy term
$U(z)$ is constructed from the spurious drift term and the pmf as

\begin{equation} \label{uze} 
  U(z) = V(z) +kT ln{R(z)} 
\end{equation}

The basic method to get the probability density and the
correlation functions is associated by solving \ref{smol2}. The
methodology lies in discretizing the equation as
\begin{equation}\label{rue} 
 \frac{d}{dt}p(i \triangle z) = a_{i-1}p((i-1) \triangle z) + b_{i+1}p((i+1) \triangle
 z)- a_{i}p(i) \triangle z) + b_{i}p((i) \triangle
 z)
\end{equation}
Monte Carlo simulations \cite{sec201} may be performed by constructing a single
barrier model, by placing two harmonic wells side by side
separated by a $6kT$ barrier. The analysis showed that two
transition rates can be obtained. With the full range of time
scales taken into account the single barrier model combine
features from both, harmonic well and discrete two state model.

\section { Dynamical Simulations}

As we have mentioned in the last section, formulating the structure-function relationships in biological ion channels has gained a new impetus with the determination of the KcsA potassium channel structure. Most of the theoretical efforts in modeling the KcsA channel have so far focused on molecular dynamics (MD) simulations of potassium ions in the channel. These studies provide valuable information on the selectivity mechanism and the energetics of ion permeation in the channel, but do not make predictions about the quantity that can be directly measured by experiment, namely the conductance. In a recent 100-ns MD simulation, \cite{sec301} calculated the conductance of a simplified channel in somewhat extreme conditions (1 M solution with a 1.1-V applied potential). This gives hope that it may be possible to determine conductance of biological channels from MD studies under physiological conditions in the not too distant future. Currently, however, typical MD simulations of biological channels can be run for ~ 10 ns, which is too short to estimate the channel conductance, or even to explore the dynamics of a single conduction event. Of course, this is not a new problem, and permeation models of lower resolution such as Brownian dynamics (BD) and Poisson-Nernst-Planck equations have long been considered in the literature. The latter approach has recently been shown to be invalid in a narrow pore environment because it neglects the self-energy of ions.

The determination of the structure of the KcsA K+ channel represents an extraordinary opportunity for understanding biological ion channels at the atomic level. In principle, molecular dynamics (MD) simulations based on detailed atomic models can complement the experimental data and help to characterize the microscopic factors that ultimately determine the permeation of ions through KcsA. A number of MD studies,\cite{sec303,sec304,sec305,sec306,sec307} broadly aimed at analyzing the dynamical motions of water molecules and ions in the KcsA channel, have now been reported. The potential functions that were used to calculate the microscopic interatomic forces and predict the the dynamical trajectory have been generated. In particular, the atomic partial charges and the Lennard-Jones radii, which are at the heart of the potential function, varied widely. Furthermore, some include all atoms (AMBER and CHARMM PARAM22), whereas others are extended-atom models that treat only the polar hydrogens able to form hydrogen bonds explicitly (CHARMM PARAM19 and GROMOS). How these differences affect the results of MD calculations is an important concern of investigation.

For meaningful theoretical studies of permeation, it is necessary to have a potential energy function providing a realistic and accurate representation of the microscopic interactions. In practice, this presents a difficult challenge. The permeation process through KcsA involves the partial dehydration of a K+ ion, followed by the translocation through the interior of a narrow pore of 12\AA{A}long, lined by backbone carbonyl oxygens, which acts as a selectivity filter\cite{sec308}. Thus, the conductance and selectivity of the KcsA channel results from a delicate balance of very strong microsopic interactions, the large energetic loss of dehydration being roughly compensated by coordination with main chain carbonyl oxygens. Gas phase experiments on model systems provide the most direct information concerning the individual microscopic interactions \cite{sec309}. High-level quantum-mechanical ab initio calculations can also be used to supplement the (often scarce) information available from experiments \cite{sec310}. The interaction of ions with a single water molecule, or with a single isolated N-methylacetamide (NMA) molecule, an excellent model of the backbone carbonyl of proteins, is of particular interest. Despite the considerable uncertainty in the experimental data and the ab initio calculations, both clearly indicate that the interaction of cations with a single NMA is substantially larger than with a single water molecule. The binding enthalpy of K+ with a water molecule is 17.9 kcal/mole, whereas it is roughly 25-30 kcal/mole with NMA. The interactions are even larger in the case of Na+. This general trend is generally reproduced by all the potential functions, with the exception of GROMOS \cite{sec311}. In this case, the interaction of K+ and Na+ with a single NMA is actually smaller than the interaction with a single water molecule. We have in our analysis have thereby relied on BD simulations with some assumptions overlaid.

BD simulations were first proposed as a way to study ion channels by \cite{sec312}. The early simulations involved one-dimensional (ID) studies of schematic channels, but the extension to three-dimensions is necessary for realistic modeling, has not been achieved until recently. The difficulty lies in the calculation of the forces on ions at each time step, typically found from the solution of Poisson's equation, which is computationally too expensive if done numerically. Thus, the first 3D BD simulations were performed by \cite{sec313,sec313a} using a torus-shaped channel, for which analytical solutions of Poisson's equation are available. For a channel with an arbitrary shape, this problem was finally resolved by storing the potential and electric field values in a set of lookup tables, and interpolating the required values during simulations from the table entries \cite{sec314}.  Recently, questions have arisen about the methods of implementing the boundaries in BD simulations of ion channels. Another problem, distinct from the issue of the boundaries, is that of accurate representation of the forces on ions in the channel. In our simulations, we plan to take into consideration of viscosity as a mean field parameter which will play a crucial role in determining the velocity profiles in the diffusion model with BD simulations giving us distinct time sclaes of the processes. Also, the calculated conductance values could be very sensitive to errors in electric fields and potentials, e.g., conductance has an exponential dependence on energy barriers in channels. A separate issue is that their treatment of forces on the ions is less sophisticated than in our simulations, in that the self-energy of the ion is ignored.

The above discussion makes it clear that some simplifying assumptions are required. In the approach described in this section, it is assumed that the protein structure is held fixed and the water molecules are replaced by a continuum. With these assumptions, the three-dimensional (3-D) movement of ion i can be described by the following simple equation:
\begin{equation}\label{br1}
m\frac{dv}{dt}= m_{i}f_{i}v_{i} + F_{R}(t)+ q_{i}E_{i}
\end{equation}
where $m_{i}, v_{i}, q_{i}$ and $f_{i}$ are the mass, velocity,
charge, and frictional coeffcient on the ith ion, respectively.
$F_{R}$ is a random thermal force representing the effects of
collisions with the water and channel wall. $E_{i}$ is the total
electrical field on the ion, including the partial charges in the
protein, all the other ions in the system, and the induced
charges from the variation in the dielectric constant at the
boundaries between the protein, water, and lipid. Although one
could always add other short-range specific force terms, this
would, in effect, be adding an empirical term that did not arise
directly from the known protein structure. The solution for this
approach proceeds as in the above molecular dynamics method. The
channel boundaries are defined, all the ions in the channel and
attached bulk reservoirs are positioned, and then, for each ion
$i$, it is integrated in discrete time steps (Brownian
dynamics). Because the dynamics of the water and protein are no
longer included and relatively long time steps can be taken for
the ion motion, this approach is many orders of magnitude faster
than the exact molecular dynamics approach (see below for a
specific example). The ability to accurately account for the
interaction between ions in the channel system is one of the most
dif?cult and critical aspects of modeling ion channels. In the
absence of such interactions, the channel conductance will vary
linearly with the ion concentration. A major advantage of this
Brownian dynamic approach is that it allows a direct simulation
of this ion-ion interaction. At each step in the dynamics, the
position of all the ions in the channel system are determined and
their interaction energy is calculated for the next time step.
One difficulty with this approach is that, because of the induced
charges at the membrane and channel water interface, the
calculation of the electrostatic energy at each step requires an
involved, time-consuming calculation. Our approach as we will
outline in subsequent sections is to ultimately play with non
equivalent role of viscosity in the outer \&\ inner pore of the
channels to modulate the conductance \&\ thereby see how the
Poisson- Boltzman equation leads to a me field dynamics with
distinct time scales.

The next simplifying approximation is to keep \ref{br1}, but
replace the exact expression for Ei by a mean field approximation
$E_{i}$ that represents a sort of average over all the possible
positions of the other ions in the system. This E is calculated
using Poisson's equation. This combination of random thermal
motion of the ion combined with a Poisson solution for E is
referred to as the Poisson-Nernst-Plank solution (PNP). The
modified 3-D steady state Nernst-Planck equation is given by:
\begin{eqnarray}\label{np1}
 0&=&\nabla\dot[\nabla c_{i}(x)+\beta \nabla V_{i}(x)c_{i}(x)] \\
V_{i}(x)&=& U(x)+z_{i}e\phi(x)+\alpha\int{\eta(x)\omega(i)}\\ \nonumber
\end{eqnarray}
where $U(x)$ is the potential due to non electrostatic forces,
$\phi$ is the electrostatic potential, $z_{i}$ is the valence of
the ith ion, and e is the electron charge and $\eta(x)$ is the
viscosity field with a weight factor $\omega(x)$ which scales
along the length of the channel.

\section{ Variable Viscosity \&\ Non Markovian Processes}

We construct our model by taking into account a large barrier
height as is obvious from our analysis in the preceding section
and taking into account the conclusions made in \cite{sec401} it is now
imperative that memory effects cannot be ignored and we use GLE
by taking into account an admixture of Non Markovian and
Markovian processes. It is noteworthy to mention here in
generalization to the PNP model that the Non Markovian process we
include here is characterize by two time scales to take into
account the features of the diffusion and drift processes in ion
channels. So in this case the generalized Langevin equation for
$n=N$ interacting ions is given by
\begin{equation}\label{gle1}
{\dot{v}_{i}} = -\int_{t_{0}}^{t}dt' \Theta(q,t-t') v(t') - V'(q)
+ \int{\omega(i)\eta(q,t-t')} + f_{i} + \lambda dW\hspace{0.5cm}
i=1 \ldots N 
\end{equation}
where f is random and systematic force acting on the ions and
$<W(t)>=0$ \&\ $<W(t_{1})W(t_{2})>=min(t_{1}, t_{2})$. The Wiener
process W is the 3-dimensional Gaussian process of which first
moment is zero vector and second moment is a diagonal matrix
whose element is minimum time between two Wiener processes. The
Brownian motion causes Wiener process that distinguishes the SDE
from the ODE, so the second term is referred as diffusion term.
Since the non-linear Eq \ref{gle1} cannot be solved analytically,
we have to integrate it numerically. The simplest numerical
method to integrate Eq \ref{gle1} is the Euler scheme. The role of
r is somewhat special, which is a parameter dependent viscous
force, where the parameter is a space dependent implying the role
of the anomalous viscous behaviour in turn linked with the kernel
as incorporated in the model. In \ref{gle1} the frictional force
depends on the previous velocities through the integral over the
kernel $\Theta(q,t-t')$, which is quantified by the fluctuation
dissipation theorem as $ <r_{\eta,i}(t')r_{\eta', j}(t)>=kT
\Theta(q,t-t')\delta_{ij}$. So in some sense we are trying to
construct a classical description of the invariant measures of 2D
Navier Stokes equation including the Stochastic effects.
Existence of  an invariant measure may be used represent the
asymptotic behavior of the system. If this invariant measure is
unique, there are chances that the law of the process solution
will converge to it. Therefore, when this holds true, this unique
invariant measure describes the equilibrium to which the system
tends. It is already known that a unique invariant measure exists
\cite{sec402} and the convergence takes place when the 2D Navier-Stokes
equations are perturbed by a time--white noise, not degenerate in
space, but with no limitations on the way it affects the modes of
the phase space. In general, without constraints on the Reynolds
number, the deterministic Navier-Stokes equations
 have many stationary solutions. No information about the long time behavior is
directly related to them. But they can be understood as invariant
  measures for the Navier-Stokes equations without the noise. Hence, our result means that, when a sufficiently distributed random perturbation
  is added, just one invariant measure exists. The effect of the noise is to mix up the dynamics of the system, allowing a unique asymptotic
  behavior.At low flow rate, the diffusion term introduces the fluctuation into the ensemble averaged stress tensor, which appears as unwanted "noise".
  This noise severely limits our ability to calculate low flow rate viscosity, where the signal to noise ratio becomes very small.
  This undesirable noise can be reduced by variance reduction method.
In simple shear flow, the velocity field is time dependent along
with the dependence on shear rate and the fluctuating viscosity.
At inception of shear flow, the system is initially at
equilibrium and the stress tensor vanishes. For time $t\geq 0$, a
constant shear rate is applied and the stresses grow until they
reach their steady state values. where the elongation rate is
indeed time dependent.

It should also be mentioned here that we make no approximations
here as regards to the relative strength of the solvent molecules
in comparison to the ions, and consequently simplification cannot
be made. The crux of the story is however is that we assume that
the memory kernel $\Theta(q,t-t')$ can be written as
\begin{equation}\label{kernel}
 \Theta(q,t-t') =  a_{0}\delta(t-t') + \frac{a_{1}}{\tau_{1}} e^{{-\frac{|t-t'|}{\tau_{1}}}} -\frac{a_{2}}{\tau_{2}} e^{{-\frac{|t-t'|}{\tau_{2}}}}
\end{equation}
which contains both Markovian \&\ Non Markovian contributions
which allows a continuous change from Markovian to Non markovian
dynamics and enables identification of terms of both the origin.
The non Markovian process has two time scales whose contributions
are dominated by the parameters $a_{1},a_{2}$ respectively. It is
also clear from the form of kernel that in the limit of weak non
Markovian process ($a_{1,2} << a_{0}$)
\begin{equation}\label{markcor}
<v_{i}(t_{0})v_{j}(t)>=kT e^{(-a_{0} t)} 
\end{equation}
with a relaxation time constant ${a_{0}}^{-1}$ which can be
determined from experimental diffusion coefficients using the
Einstien relation. Similarly in the limit of weak Markovian noise
the spectral density can be evaluated as
\begin{equation}\label{spec}
S(\omega) =
\frac{2kT\omega^{2}}{(1+\tau_{1}^2\omega^{2})(1+\tau_{2}^2\omega^{2})}
\end{equation}
We get the spectrum limits for $ \tau_{1,2}\rightarrow 0$ which
give us combination of the color noise. In a similar tone provided we fix the kernel by taking into consideration the velocity profile we will be left with a viscosity term modulated in two time scales,

\begin{equation}\label{kernel1}
\eta((q),t-t')= {{\alpha_{1}{(\tau'_{1})} e^{{-\frac{|t-t'|}{\tau'_{1}}}}} -{{\alpha_{2}{(\tau'_{2})}}e^{{-\frac{|t-t'|}{\tau'_{2}}}}}}\\ \nonumber 
\end{equation}

So physically the inputs
for these simulations has been to assume that water inside the
pore is different from that at the mouth. So inside the channel
the gluelike properties of water with a shear viscosity has the
capabilities to give rise to Nonmarkovian processes with memory
and the thereby the whole gating processes need to be
investigated.

The most challenging aspect of simulation of ion channels has
been the implementation of particle coupling and boundary
conditions. In BD simulations it is particularly difficult to
maintain the correct particle concentrations and behaviour that
occur under extreme conditions that occur in channels
simulations. Here we use a simple algorithm in order to solve, the
GlE can be written as a set of 3 LE's
\begin{eqnarray}\label{le}
  \ddot{z} &=& -a_{0} V'(z) + \beta_{1} + \beta_{2}   \nonumber \\
  \dot{\beta_{1,2}} &=& -\frac{-\beta_{1,2}}{\tau_{1,2}}-
  \frac{a_{1,2}}{\tau_{1,2}}\dot{z}+ r_{i} + f_{i}
\end{eqnarray}
The distrubution functions from our simulations are shown below which show a specific correlation.
  
We can use the second order Runge-Kutta method by discretizing the
equation \ref{le} and approximating the potential as $
\int_{t}^{t+\triangle t} V(x(s))ds = V(x(t)\triangle t$ We need
$n$ Gaussian random numbers to be picked at each step for the
algorithm. Integrations of noise can be simulated by linear
combinations of 3 normal Gaussian random numbers and the matrix
elements of the coefficients are evaluated by the self and cross
corelations of the integrations of noise. It should be mentioned
here that some interesting results have been obtained by
considering a Gillepsie Algorithm in the context of BD
simulation, which has been not considered here.

\section{Results}

As discussed in the previous section we apply the algorithm and
get a comparison between the time scales of diffusion and drift
performed in C++. As we have shown previously the model has the
ability to describe two time scales behaviour along with
oscillatory behaviour where the velocity autocorrelations exhibit
an exponential decay. The Brownian simulation for GLE has been
performed by subdividing the charge of the particle and
increasing the number of particles by the same factor. We have
obtained the velocity distributions corresponding to different
limits of the processes and is explicit that oscillations
dominate at the limit of weak Non Markovian process. We have
shown here the figures of a Non Markovian process that gives rise
to a memory function along with the two time scales in the BD
behaviour and also we find out the standard deviations of the
results.

The results of the simulations clearly show that BD approach can
be used to describe Coulomb interaction and different time scale
phenomena. It is also clear that single barrier model with a
threshold for the barrier plays an important role for charge
transport through ion channels and advocates use of particle
based simulation models. But the most striking feature of the
model is that we may be able to model both the memory effects of
gating and the oscillatory behaviour by a GLE which contains an
admixture of Markovian \&\ Non Markovian processes.

The admixture of the time evolution kernel \&\ the viscosity
terms makes it clear and the correlation function makes it clear
a phase transition of the shear viscousity within the channels may give an important
clue as regards the rise of memory effects and thereby the evoution of flip flop motion within Potassium Channels.
The exact Stochastic Simulation using Non Linear 3D Navier needs to be shown.

\newpage
\begin{figure}
\epsfxsize=4in \centerline{\epsfbox{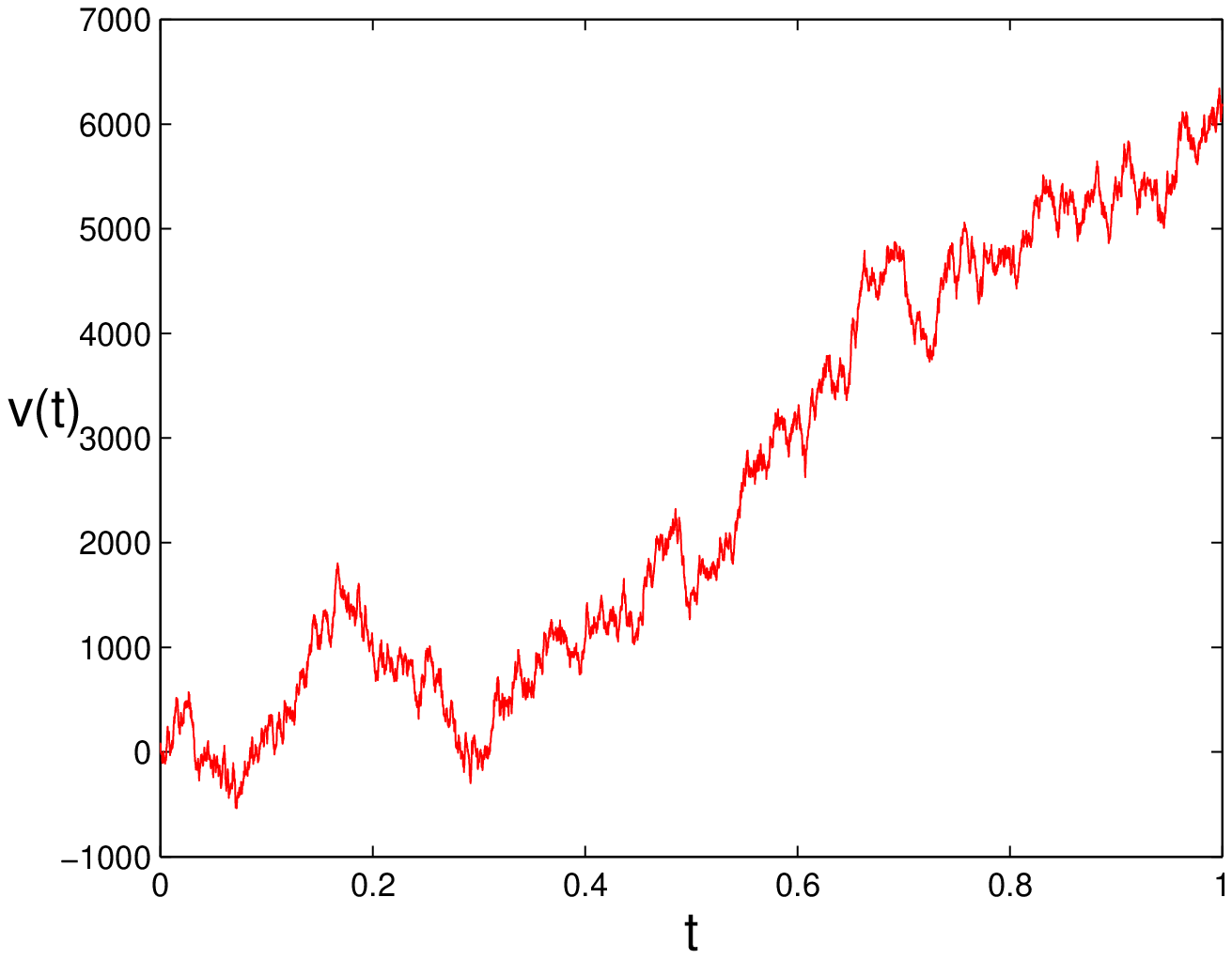}}
\label*{}\caption {Velocity Plots with no correlation}
\end{figure}

\begin{figure}
\epsfxsize=4in \centerline{\epsfbox{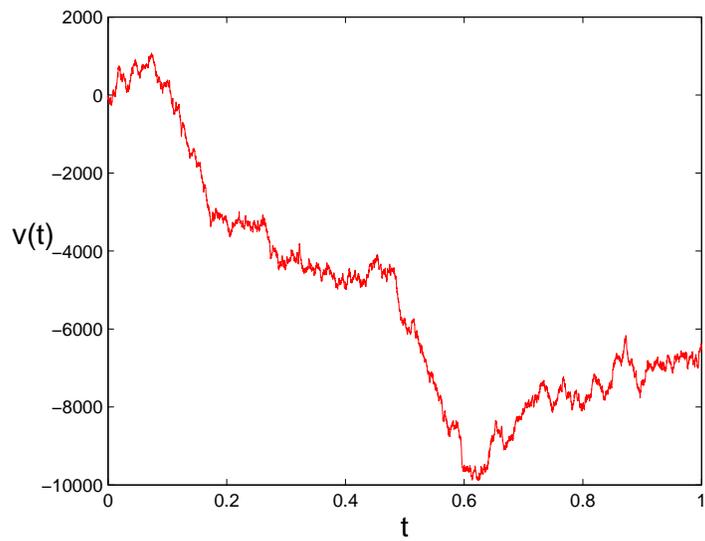}}
\label*{}\caption {Velocity Plots at length scales where viscosity is weak}
\end{figure}
\newpage
\begin{figure}
\epsfxsize=4in \centerline{\epsfbox{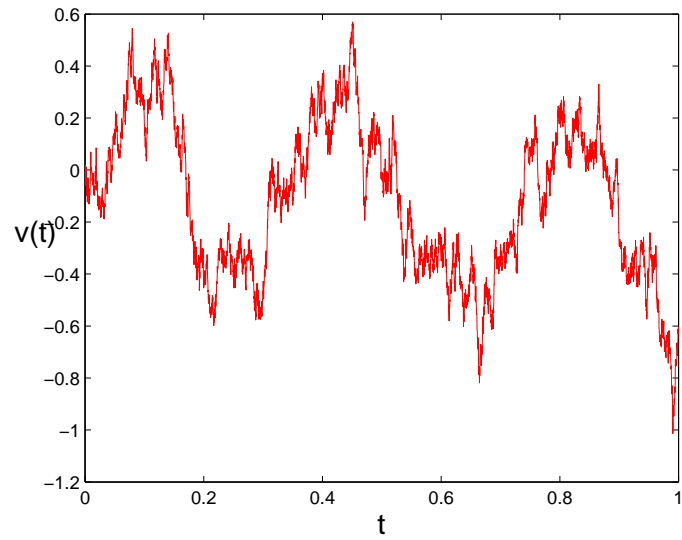}}
\label*{}\caption {Velocity Plots at length scales where viscosity is dominant}
\end{figure}

\begin{figure}
\epsfxsize=4in \centerline{\epsfbox{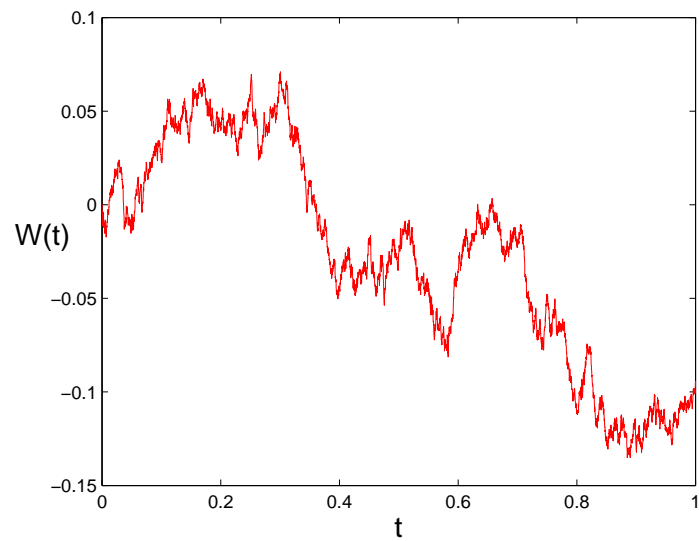}}
\label*{}\caption {Plots of Velocity profiles from Simulations which indicate the role of Viscosity at different Time Scales}
\end{figure}

\begin{figure}
\epsfxsize=4in \centerline{\epsfbox{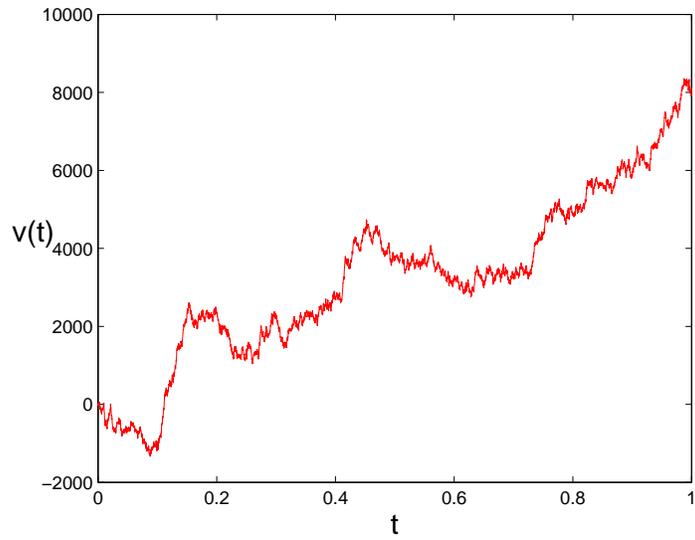}}
\label*{}\caption {Velocity plots with no correlation upto 10000 iteraions}
\end{figure}
\begin{figure}
\epsfxsize=4in \centerline{\epsfbox{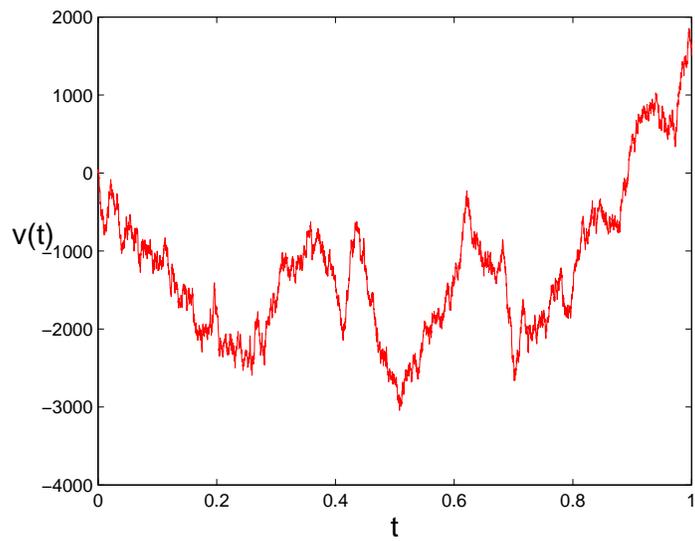}}
\label*{}\caption {Velocity plots in weak non-markovian limit}
\end{figure}

\end{document}